\documentstyle[prl,aps,multicols,epsf]{revtex}
\def\beginwide{
        \end{multicols} \vspace*{-0.5cm} \noindent
        \rule{3.5in}{.1mm}\rule{.1mm}{5mm} \widetext \medskip }
\def\endwide{
        \hspace*{3.5in}~\rule[-5mm]{.1mm}{5mm}\rule{3.5in}{.1mm}
        \begin{multicols}{2}\narrowtext \vspace*{-1.0cm} \noindent }
\def\beginwidetop{
        \end{multicols} \vspace*{-0.5cm} \noindent
        \widetext \medskip }
\def\endwidebottom{
        \begin{multicols}{2} \vspace*{-1.0cm} \noindent }

\begin{document}
\draft
\title{Scalar Fields as Dark Matter in Spiral Galaxies}  

\author{ F. Siddhartha Guzm\'an\footnote{E-mail: siddh@fis.cinvestav.mx}
and Tonatiuh Matos\footnote{E-mail: tmatos@fis.cinvestav.mx}}
\address{Departamento de F\'{\i}sica,\\
Centro de Investigaci\'on y de Estudios Avanzados del IPN,\\
AP 14-740, 07000 M\'exico D.F., MEXICO.\\}

\date{\today}
\maketitle

\begin{abstract}
We present a model for the dark matter in spiral galaxies, 
which is a result of a static and axial symmetric exact solution of the 
Einstein-Dilaton theory. We suposse that dark matter is a scalar field 
endowed with a scalar potential. We obtain that a) the effective energy 
density goes like $1/(r^2+r_{c}^{2})$ and b) the resulting circular 
velocity profile of tests particles is in good agreement with the
observed one.
\end{abstract}

\begin{multicols}{2}
\narrowtext

One of the greatest puzzles of physics at the moment is without doubt the
existence of dark matter in cosmos. The experimental fact that the galaxy
masses measured with dynamical methods do not coincide with their luminous
galaxy masses gives rise to the existence of a great amount of dark matter
in galaxies, galaxy clusters and superclusters. At the present time,
cosmological observations indicate that the universe is filled out with
about 90 percent of dark matter, whose nature till now remains
unexplained. Recently some authors have proposed the scalar field as a
candidate for dark matter in cosmos \cite{cho,dick}, in some sense the
inflationary cosmological model proposes the scalar field as cosmological
dark matter as well. These models consider scalar-tensor theories of
gravity where one is able to add mass terms to the total energy density of
the space-time. All modern unifying field theories also contain scalar
fields. For example, scalar fields are fundamental fields in Kaluza-Klein
and Superstring theories, because such fields appear in a natural way
after dimensional reduction. In both theories the scalar field could be
endowed with an exponential scalar potential \cite{frolov,merced}, in
particular, when we deal with 5-dimensional Kaluza-Klein theories, the
Lagrangian density reads ${\cal L}_5 = R_5 + \Lambda$ being $\Lambda$ a
5-dimensional cosmological constant. After dimensional reduction and a
conformal transformation one obtains the density ${\cal L}_4 = -R_4 +
2(\nabla \phi)^2 +e^{-2/\sqrt{3}\phi}\Lambda$, where $\phi$ is the scalar
field which actually states that an exponential potential appears in a
natural way in this theory. An analogous procedure stablishes that in the
low energy limit of Superstring theory one gets a similar result
\cite{cho,frolov}. In general one obtains the Lagrangian from
high-dimensional theories ${\cal L}_4 = -R_4 + 2(\nabla \Phi)^2 +
e^{-2\alpha\Phi}\Lambda$, therefore here we will restrict our selves to an
exponential scalar potential. In this letter we show a possible model for
the dark matter in spiral galaxies, supposing that such matter is of
scalar nature.\newline

There is a common approach to explain the rotation curves in spiral
galaxies called Modified Newtonian Dynamics (MOND) \cite{milgrom,begeman},
which basically consists of modifying the Newton's law of attraction for
small accelerations by adding terms to the gravitational potential. In
this way, by adjusting some free parameters for each galaxy, one can
reproduce the asymptotic behavior of the rotation curves. However it
appears to be artificial because it is nothing but a mere correction of
Newton's law, we are unable to know neither where the parameters and the
correction terms come from, nor why nature behaves like that and therefore
which is the Newton's law at cosmological scale for instance.\newline

A convincing phenomenological model for galactic dark matter is the called
Isothermal Halo Model (IHM), which assumes the dark matter to be a
self-gravitating ball of ideal gas (made of any kind of particles) at a
uniform temperature $kT=\frac{1}{2}m_{dm}v_c$, - being $m_{dm}$ the mass
of each particle and $v_c$ its velocity - which eventually produces a dark
matter distribution going as $\sim 1/r^2$, implying in this way an
increasing mass $M(r) \sim r$. Then, by assuming that a galaxy is a system
in equilibrium ($GM/r^2 = v_c^2/r$) the velocity of particles surrounding
the profile above should produce flat rotation curves into a region where
the dark matter dominates, i.e. at large radii when one considers as
exponential distribution of luminous matter as usual
\cite{peebles}.\newline

Observational data show that galaxies are composed by almost 90\% of
dark matter \cite{begeman,peebles,persic}. This is so because the
kinematics
inside the dark matter dominated region is not consistent with the
predictions of Newtonian theory, which explains well the dynamics of the
luminous sector of the galaxy but predicts a keplerian falling off for
the rotation curve. The region of the galaxy we are interested
in is that in which the dark matter determines the kinematics of test
particles. So we can suppose that luminous matter does not contribute
in a very important way to the total energy density of the matter that 
determines the bahavior of particles in the mentioned region, instead the
scalar matter will be the main contributor to it. Thus, as a first
approximation we can neglect the baryonic matter contribution to the total
energy  density for the explanation of assimptotic rotation 
curves.\newline

On the other hand, the exact symmetry of the dark halo stills unknown, but
it is reasonable to suppose that it is symmetric with respect to the
rotation axis of the galaxy. In this letter we let the symmetry of the
halo as general as we can, so we choose it to be axial symmetric.
Furthermore, the rotation of the galaxy does not affect the motion of
test particles around the galaxy, dragging effects in the halo of the 
galaxy should be too small to affect the tests particles (stars) traveling 
around the galaxy. Hence, in the region of our interest we can suppose the 
space-time to be static, given that the circular velocity of stars (like
the sun) of about 230 Km/s seems not to be affected by the rotation of the 
galaxy and we can consider a time reversal symmetry of the space-time.
So, the model we are dealing with will be given by the gravitational 
interaction modified by a scalar field and a scalar potential. The model 
consists of the following action

\begin{equation}
S = \int d^4x\sqrt{-g}[-\frac{R}{\kappa_0} + 2(\nabla \Phi)^2 - V(\Phi)],
\label{1}
\end{equation}
which could be the four-dimensional action for the Kaluza-Klein or the Low 
Energy Superstings theory without electromagnetic
field, and where we have added a term which contains the scalar potential. In
this action $R$ is the scalar curvature, $\Phi$ is the scalar field, 
$\kappa_0 = \frac{16\pi G}{c^3}$ and $\sqrt{-g}$ is the determinant of the 
metric. The most general static and axial symmetric line element 
compatible with this action, written in the Papapetrou form is 

\begin{equation}
ds^2 = \frac{1}{f}[e^{2k}(dzd\bar{z}) + W^2d\phi^2] - f\ c^2dt^2,
\label{2}
\end{equation}
where $z:=\rho + i\zeta$ and $\bar{z} :=\rho
-i\zeta$ and the functions $f,\ W$ and $k$ depend only on $\rho$ and $\zeta$. 
This metric represents the symmetries posted above.
The application of the variational principle to (\ref{1}) gives rise to the
field equations
\begin{eqnarray}
&\Phi^{;\mu}_{;\mu} + \frac{1}{4} \frac{dV}{d\Phi} = 0 \cr\cr
&R_{\mu \nu} = \kappa_0 [2 \Phi_{,\mu} \Phi_{,\nu} - \frac{1}{2} g_{\mu
\nu} V(\Phi)],
\label{3}
\end{eqnarray}
which are the Klein-Gordon and Einstein's field equations respectively; 
$\mu,\nu=0,1,2,3$. Using the harmonic maps ansatz
\cite{matos1,matos2,matos3}
we find the 
following Poisson like structure for the above equations \cite{siddh}
\begin{eqnarray}
\hat{\Delta}\lambda &=& - \kappa_0 \sqrt{-g} V(\Phi) \cr\cr
2\hat{\Delta} \Phi &=& \frac{1}{4}\sqrt{-g}\frac{dV}{d\Phi} \cr\cr
\label{4}
W_{,z\bar{z}} &=& -\frac{1}{2} \kappa_0 \sqrt{-g}V(\Phi) \cr\cr
k_{,z} &=& \frac{W_{,zz}}{2W_{,z}} + \frac{1}{4}W\ {\lambda_{,z}}^{2}W_{,z} +
\kappa_0W {\Phi_{,z}}^{2}W_{,z},
\end{eqnarray}
and a similar equation for $k_{,\bar{z}}$, with $\bar{z}$ instead of $z$,
where 
$\hat{\Delta}$ is the Laplace operator such that for any function
$h=h(z,\bar{z})$: $\hat{\Delta}h := (Wh_{,z})_{,\bar{z}} +
(Wh_{,\bar{z}})_{,z}$. Moreover, $\lambda=\ln(f)$ is interpreted as the
gravitational potential.\newline

If one assumes that $\lambda $ and $\Phi $ depend only on $W(z,\bar{z})$,
the set of equations (\ref{4}) appears in a more tractable
form

\begin{eqnarray}
2WW_{,z\bar{z}}D\lambda  &=&-\kappa _{0}\sqrt{-g}V(\Phi )  \label{lambda}
\\
2WW_{,z\bar{z}}D\Phi  &=&\frac{1}{4}\sqrt{-g}\frac{dV}{d\Phi }
\label{Phi} 
\\
W_{,z\bar{z}} &=&-\frac{1}{2}\kappa _{0}\sqrt{-g}V(\Phi )  \label{W} \\
k_{,z} &=&\frac{W_{,zz}}{2W_{,z}}+\frac{1}{4}W\ {\lambda ^{\prime }}%
^{2}W_{,z}+\kappa _{0}W{\Phi ^{\prime }}^{2}W_{,z}  \label{k}
\end{eqnarray}  
and the corresponding expression for $k_{\bar{z}}$, where now the operator
$D$ means $Dh(W)=Wh^{\prime \prime }+2h^{\prime }$ $\forall h=h(W)$, and $%
^{\prime }$ denotes derivative with respect to $W$. Equations
(\ref{lambda}- \ref{k}) constitute a system of coupled differential
equations because $\sqrt{-g}=We^{2k-\lambda }/2$. However it is evident
that once we have expressions for $\lambda $ and $\Phi $, $k$ can be
integrated by quadratures. Moreover, using the third of these equations
$\lambda $ and $\Phi $ obey differential equations with $W$ being the
independent variable.\newline

In order to find an exact solution, we substitute $\kappa
_{0}\sqrt{-g}V(\Phi )$ from (\ref{W}) into (\ref{lambda}) and (\ref{Phi}),
(remember that $\frac{dV} {d\Phi }=-2\alpha V$) and obtain two decoupled
differential equations, one for $\lambda $ and another for $\Phi .$ We
solve these two differential equations and substitute the solution into
(\ref{k}). We thus find that a solution of the system
(\ref{lambda}-\ref{k}) is given by
  
\begin{eqnarray}
\lambda  &=&\ln (M)+\ln (f_{0})  \nonumber \\
\Phi  &=&\Phi _{0}+\frac{1}{2\sqrt{\kappa _{0}}}\ln (M)  \nonumber \\
V &=&\frac{4f_{0}}{\kappa _{0}M}  \nonumber \\
k &=&\frac{1}{2}(\ln {M}_{,z\bar{z}}+\ln {M})  \label{sol}
\end{eqnarray}

\noindent
where $f_0$ and $\Phi_0$ are integration constants and $W=M$ is a function
restricted only by the condition
\begin{equation}
MM_{,z\bar{z}} = M_{,z}M_{,\bar{z}} 
\label{M}
\end{equation}
whose solutions are $M=Z(z)\bar{Z}(\bar{z}),$ where $Z$ is an arbitrary
function. The reader can check that (\ref{sol}) is a solution of the field
equations substituting the set (\ref{sol}) into (\ref{3}) using the
metric (\ref{2}).\newline

In what follows we study the circular trajectories of a test particle on the
equatorial plane taking the space-time (\ref{2}) as the background. The 
motion equation of a test particle in the space-time (\ref{2}) can be derived 
from the Lagrangian

\[
{\cal L}=\frac{1}{f}[e^{2k}(\left(\frac{d\rho}{d\tau}\right)^2+
\left(\frac{d\zeta}{d\tau}\right)^2) +
W^2\left(\frac{d\phi}{d\tau}\right)^2] 
\]
\begin{equation}
-f\ c^2\left(\frac{d\ t}{d\tau}\right)^2.
\label{lgeo}
\end{equation}

This Lagrangian contains two constants of motion, the angular momentum 
per unit of mass
\begin{equation}
\frac{W^2}{f}\frac{d\phi}{d\tau}=B,
\label{B}
\end{equation}
and the total energy per unit of mass of the test particle
\begin{equation}
f\ c^2\frac{d\ t}{d\tau}=A,
\label{A}
\end{equation}
where $\tau$ is the proper time of the test particle. An observer falling 
freely into the galaxy, with coordinates $(\rho,\zeta,\phi,t)$, will have a 
line element given by
\begin{eqnarray}
ds^2 &=& \left\{\frac{1}{f\ c^2}[e^{2k}({\dot\rho}^2+{\dot\zeta}^2) + 
W^2{\dot\phi}^2] - f\right\} c^2dt^2\cr\cr
&=& \left(\frac{v^2}{c^2}-f \right)c^2dt^2\cr\cr
&=& -c^2d\tau^2.
\label{line}
\end{eqnarray}
The velocity $v^a=(\dot{\rho},\dot\zeta,\dot{\phi})$, is the
three-velocity of the test particle, where a dot means derivative with
respect to $t$, the time measured by the free falling observer. The squared 
velocity $v^2$ is then
\begin{equation}
v^2=g_{ab}v^av^b = \frac{e^{2k}}{f}(\dot{\rho}^2 +
\dot{\zeta}^2)+\frac{W^2}{f}\dot{\phi}^2,
\label{vel}
\end{equation}
where $a,b=1,2,3$. Using (\ref{line}) into (\ref{A}) we obtain an 
expression for the squared energy
\begin{equation}
A^2=\frac{c^4 f^2}{f-\frac{v^2}{c^2}}.
\label{Ac}
\end{equation}
We are interested in test particles (stars) moving on the equatorial plane 
$\dot\zeta=0$ and the equation of motion derived from the geodesics of
metric (\ref{2}) reads
\begin{equation}
\frac{1}{f}e^{2k}\left(\frac{d\rho}{d\tau}\right)^2 + \frac{B^2\ f}{W^2} - 
\frac{A^2}{c^2f}=-c^2.
\label{egeo}
\end{equation}
where we have used the conservation equations (\ref{B}) and (\ref{A}). 
Equation (\ref{egeo}) determines the trajectory of a test particle around the 
equator of the galaxy, in this trajectory $A$ and $B$ remain constant. If we 
change of test particle, we could have another constants of motion $A$ and
$B$ determining the trajectory of the new particle. A spiral galaxy is
practically a disc of stars traveling around the equatorial plane of the
galaxy in circular trajectories in the period of observation, altough it
had to be formed from enormous clouds of gas going around a symmetry axis 
with average values of $A$ and $B$.
Thus for a circular trajectory $\dot\rho=0$, the equation of motion 
transforms into
\begin{equation}
\frac{B^2\ f}{W^2} - \frac{A^2}{c^2f}=-c^2.
\label{ecgeo}
\end{equation}
This last equation determines the circular trajectories of test particles 
travelling on the equator of the galaxy. Using (\ref{ecgeo}) and
(\ref{Ac}) we 
find an expression for $B$ in terms of $v^2$,
\begin{eqnarray}
B^2&=&\frac{v^2}{f-\frac{v^2}{c^2}}\frac{W^2}{f}\cr\cr\cr
&\sim& v^2\frac{W^2}{f^2},
\label{Bc}
\end{eqnarray}
since $v^2 \ll c^2$. Now using (\ref{Bc}) one concludes that for our
solution (\ref{sol}) $v^2=f_{0}^{2}B^2$, $i.e.$
\begin{equation}
v_{DM}=f_0B,
\label{Ma} 
\end{equation}
where we call $v\rightarrow v_{DM}$ the contribution of our dark matter
to the circular velocity of a star.\newline

When $Z=z$ our solution in 
Boyer-Lindquist coordinates $\rho=\sqrt{r^2-2ar+b^2}\sin\theta$,
$\zeta=(r-a)\cos\theta$ reads

\[
ds^2 = \frac{(1-\frac{a}{r})^2 + \frac{K^2\cos^2\theta}{r^2}}{f_0
r_0}(\frac{dr^2}{1-2\frac{a}{r}+\frac{b^2}{r^2}} +
r^2\ d\theta^2)
\]
\[
+\frac{(r-a)^2 + K^2\sin^2\theta}{f_0 r_0} d\phi^2 
\]
\begin{equation}
-f_0 c^2\frac{(r-a)^2 + K^2\sin^2\theta}{r_0} dt^2
\label{DM}
\end{equation}
where $K^2=b^2-a^2$ and $r_0$ only scales. 
The effective energy density $\mu_{DM}$ of (\ref{sol}) is given by
the expression

\begin{equation}
\mu_{DM} = \frac{1}{2}V(\Phi) = \frac{2f_0r_0}{\kappa_0 ((r-a)^2 +
K^2\sin^2\theta)}
\label{muDM}
\end{equation}

\noindent
and plays the role of our dark matter density profile.\newline

Keeping in mind that this is only the contribution of dark matter to the
energy density we are in conditions to compare these results with those
given by measurements. In order to do so we recall the paper by
Begeman et al. \cite{begeman} where an energy
density profile of the IHM $\mu(r) = \rho_0 r_{c}^2/(r^2+ r_{c}^{2})$
for dark matter is used, being $r_c$ a core radius.
It is evident that this profile is a particular case of the expression we
present here, namely, for matter localized on the equator of the galaxy. 
So, we can fit some of the free parameters of metric (\ref{DM}) 
comparing these two profiles. We set $b=r_c,\ a=0$ and 
$\frac{2f_0r_0}{\kappa_0}=\rho_0r_c^2$.\newline

\begin{figure*}
\label{fig1}
\leftline{ \epsfxsize=4cm \epsfbox{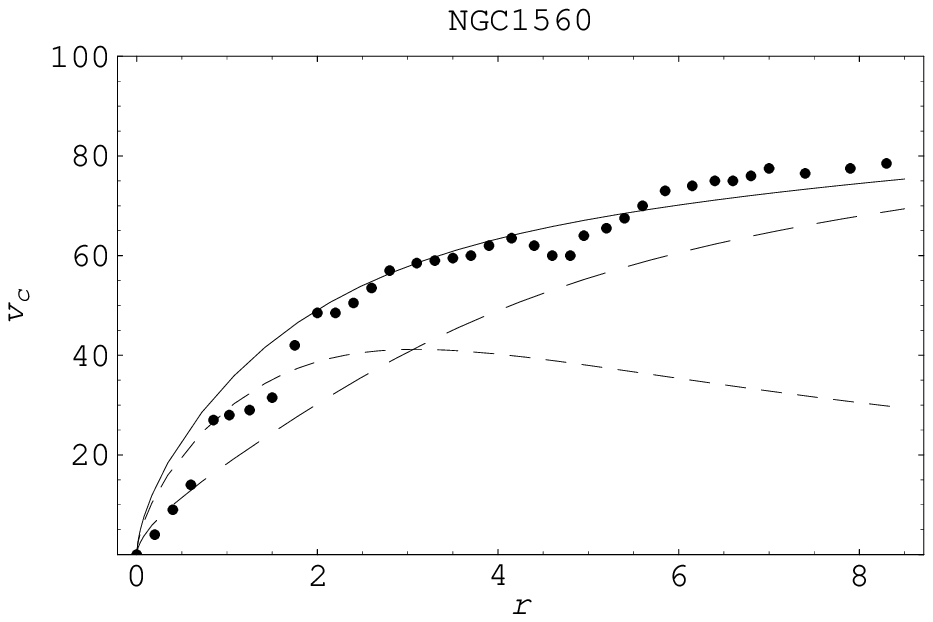}
\epsfxsize=4cm \epsfbox{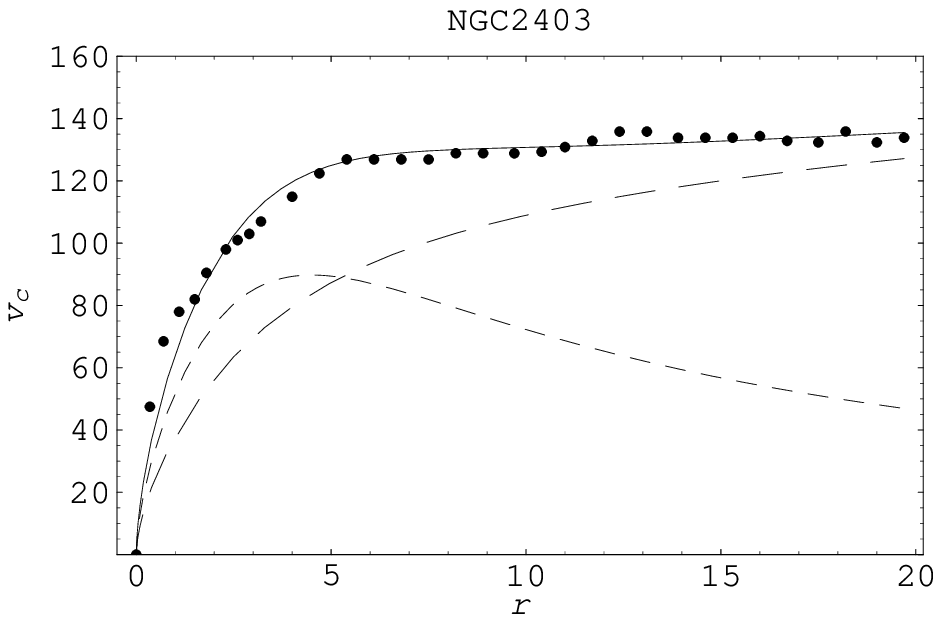}}
\leftline{ \epsfxsize=4cm \epsfbox{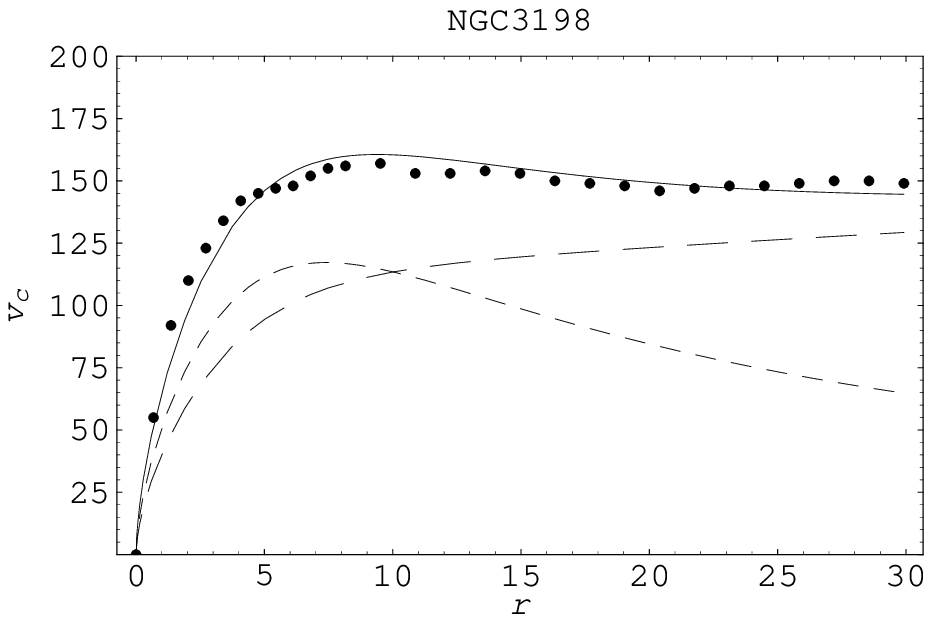} \epsfxsize=4cm
\epsfbox{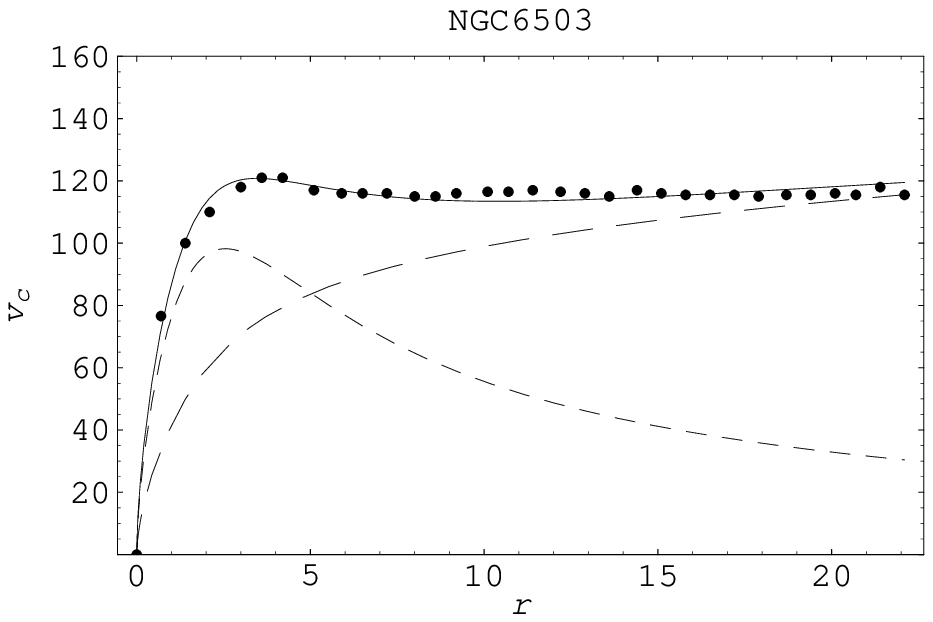}}
\caption{The circular velocity profiles of four spiral galaxies.
Solid lines represent the total circular velocity ($v_C$), 
long-dashed is the contribution of the dark matter to the total velocity
($v_{DM}$) and the short-dashed curves is the contribution of luminous
matter ($v_L$); finally the dots represent the observational data. The
units are in (Km/s) in the vertical-axis and in (Kpc) in the
horizontal-axis.} 
\end{figure*}

Let us model the circular velocity profile due to the luminous matter
of the disc in a spiral galaxy by the function
\begin{equation}
v_{L}^{2} = v^2(R_{opt})\beta \frac{1.97 x^{1.22}}{(x^2 + 0.78^2)^{1.43}}
\label{vL}
\end{equation}
which is the aproximatted model for the Universal Rotation Curves (URC) as
was propossed by Persic et al. \cite{persic} for an exponential thin disc,
valid for a sample of 967 galaxies; in this expression $x=r/R_{opt}$, the
parameter $\beta=v_{L}(R_{opt})/v(R_{opt})$ being $R_{opt}$ the radius 
into which it is contained the 83\% of the onservable mass of the galaxy
and $v$ the observed circular velocity.\newline

We can suppose that luminous matter near the center of a galaxy behaves
like in Newtonian mechanics. Thus with the luminous velocity (\ref{vL}) it
is now easy to calculate the angular momentum (per unity of mass) of the
test particle in the luminous matter dominated region
\begin{equation}
B=v_LD,
\label{vLD}
\end{equation}
where $D$ is the distance between the center of the galaxy and the test 
particle. For our metric, $D=\int ds$, keeping $\theta,\ \phi$ and $t$
constant, we obtain $D=\sqrt{(r^2-2ar+b^2)/f_0 r_0}$. Observe that after
we have determined the dark matter energy density $\mu_{DM},\ B$ is
uniquely determined by $v_L$ via (\ref{vLD}); it is easy to show that $B$
in (\ref{vLD}) equals that of equations (\ref{egeo}-\ref{Ma}) by
including a luminous
newtonian component into the radial geodesic equation \cite{dario}.
Therefore (\ref{Ma}) and (\ref{vLD}) imply the total circular velocity 

\begin{equation}
v_C = \sqrt{v_{L}^{2} + v_{DM}^{2}} = v_L \sqrt{1 + f_0 (r^2 - 2ar + b^2)}
\label{vtot}
\end{equation}

\noindent
expression that should fit the observed rotation curves. In order to do
so, we present in Fig. 1 the plots containing the fittings of four spiral
galaxies and into Table \ref{tab1} the values of the parameters
$f_0$ and $b$ keeping
$a=0$ and the scale $r_0=1$. From it we see that the agreement of the
resulting circular velocity profiles given by the scalar field as dark
matter and the observed is very good not only far away from the center of
the galaxy but inside the part of the galaxy dominated by luminous matter
as well.\newline

The criterion used to choose the sample was the ratio of the dark to
luminous mass inside $r_{25}$, which was
selected to be $\sim 1$ in order to test our model by using "dark
enough galaxies". The plots shown in Fig. 1 would not be enough to
state that our model works, it is necessary to be consistent with the
phenomenological URC approach, i.e. the contribution of our dark matter
should be the same as that propossed into the URC frame which is strongly
luminosity dependent. A formula consistent with (\ref{vL}) is given by
\cite{persic}

\begin{equation}
v^{2}_{urc DM} = v^2(R_{opt})(1-\beta)(1+\gamma^2)\frac{x^2}{x^2+\gamma
^2}
\label{urcdm}
\end{equation}

\noindent
being $\beta = 0.72 + 0.44\log L/L_{*}$ the same parameter as in
(\ref{vL}) and $\gamma =1.5 (L/L_{*})^{1/5}$.\\

According to (\ref{Ma}) and (\ref{vLD}) the contribution of our dark
matter
is

\begin{equation}
v^{2}_{DM} = f_0(r^2+b^2)v^2(R_{opt})\beta
\frac{1.97x^{1.22}}{(x^2 + 0.78^2)^{1.43}}
\label{ourdm}
\end{equation}

\noindent
which after using the fitting parameters of Table \ref{tab1} the
comparisson of both approaches are compared in Fig. 2 from
which it can be concluded that our dark matter model respects
the luminous matter model we have used.\\

\begin{figure*}
\label{fig2}
\leftline{ \epsfxsize=4cm \epsfbox{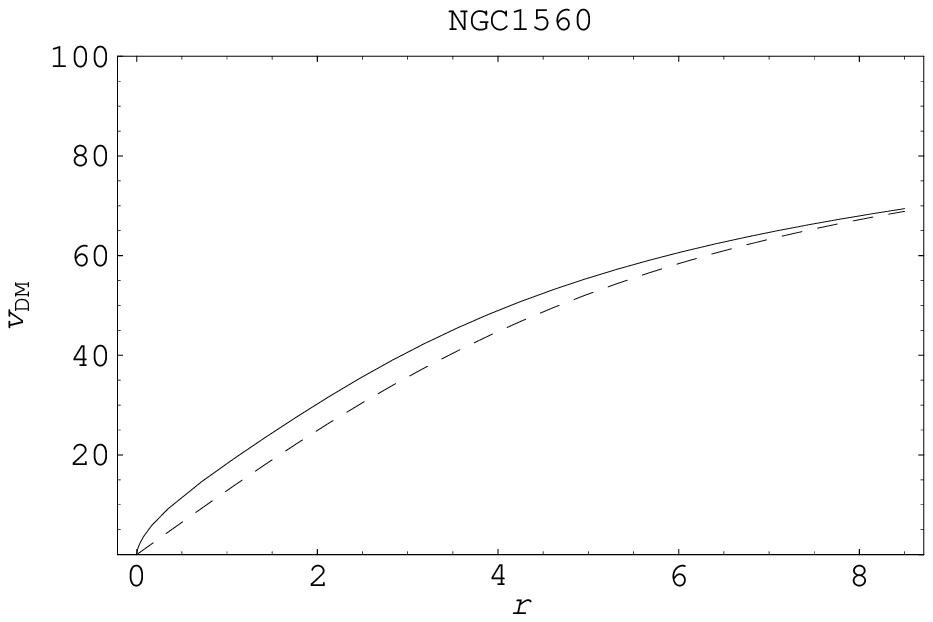}
\epsfxsize=4cm \epsfbox{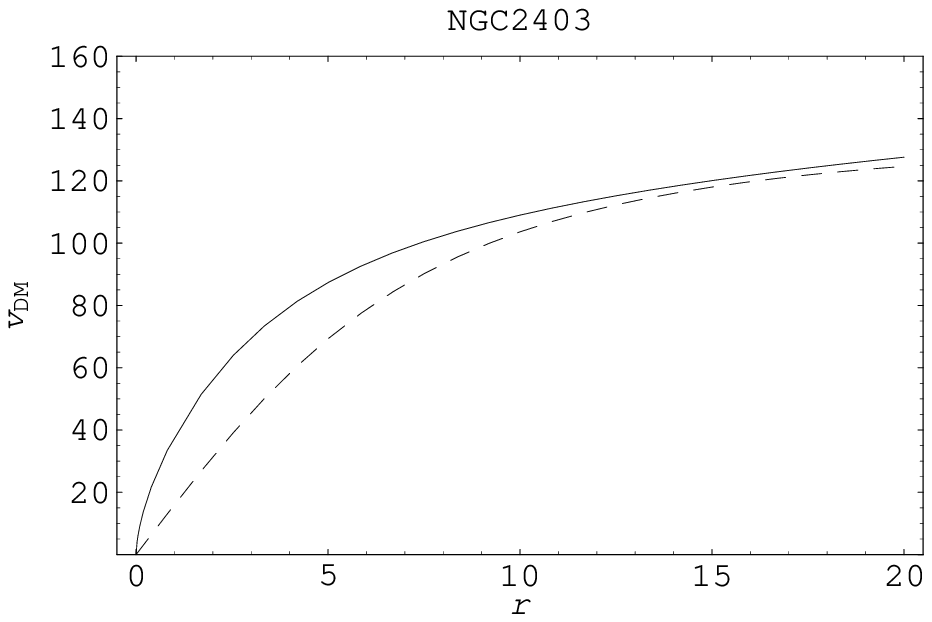}}
\leftline{ \epsfxsize=4cm \epsfbox{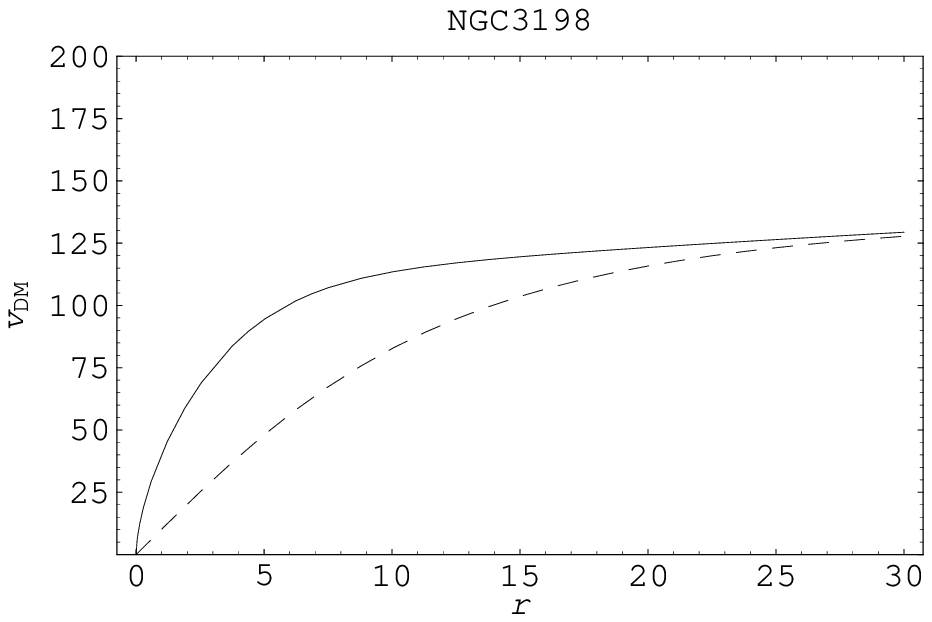} \epsfxsize=4cm
\epsfbox{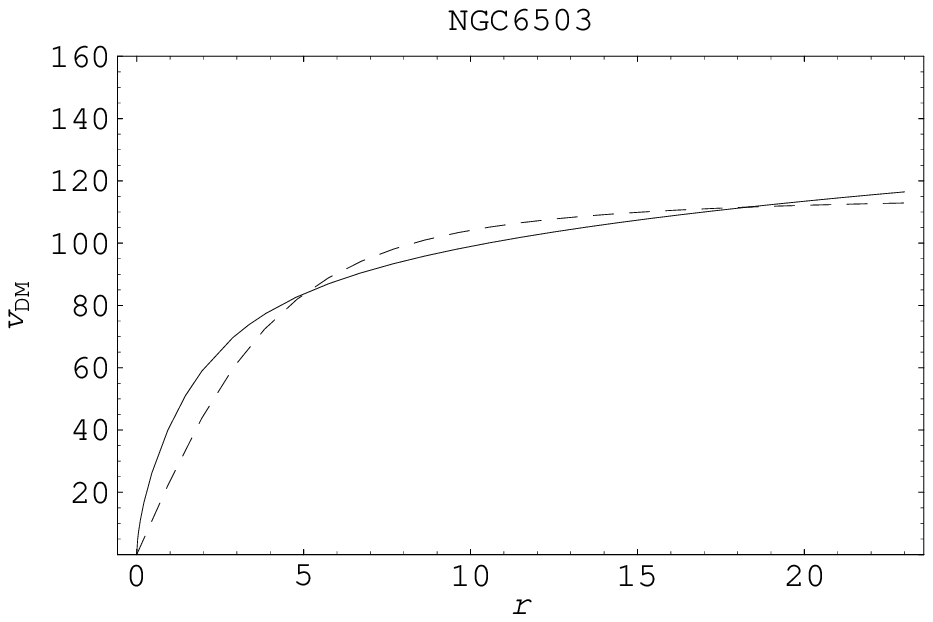}}
\caption{The contribution of dark matter to the circular velocity of
test particles is shown. Solid lines emerge from the model described in
this letter, dashed ones correspond to the URC approach. The discrepancies 
are of 8.3, 11, 23.3 and 5.2 percent respectively.} 
\end{figure*}

Some remarks can be drawn. The energy density (\ref{muDM}) coincides 
with that required for a galaxy to explain the rotation curves of test
particles in its halo, but in our model, this energy density is product
of the scalar field and the scalar field potential, that is, this dark
matter is produced by a $\Phi$ particle. So we have shown that there is
an exact solution which describes the rotation curves of
particles in a spiral galaxy. The crucial point for having the circular 
velocity $v_{DM}=f_0B$ is that $f\sim W$ in the solution (\ref{sol}). 
But this fact remains unaltered after conformal transformations in the
metric $d\hat{s}^2=A(\Phi)ds^2$, so that the circular velocity $v_{DM}$
remains the same for all theories and frames related with metric
(\ref{2}) by conformal transformations.\newline

How does our model looks like into the cosmological context? 
When a density profile for galactic dark matter goes as the inverse of
$r^2$ and it is supossed that the halo of a galaxy ends in the region
where those of neighboring galaxies start, the integrated amount of
galactic dark matter is close to that needed for the Universe to be flat
for the observed average distance between them \cite{peebles}, flatness
inferred from the cosmic background radiation \cite{riess} and thus 
permiting our model to be inside the bounds. In fact, we have
developed a cosmological model that considers the same theory as here
(\ref{1}) with the same scalar potential \cite{luis}, which has been able
to explain the redshifts of Type Ia supernovea, and all the parameters for
structure formation lie into the ranges impossed by observations
\cite{riess,stein}, which make us to put forward the model presented in
this letter.\newline

\acknowledgements{
We want to thank Ulises Nucamendi and Hugo Villegas Brena for many 
helpful discussions. This work was partly supported by CONACyT,
M\'exico, under grants 94890 (Guzm\'an) and 3697-E (Matos).}

\begin{table}[h]
\caption{Values of the fitting parameters $f_0$ and $b$. Here also are
shown the values of the quantities used.}
\label{tab1}
\begin{center}
\begin{tabular}{ccccc}\hline\hline
Galaxy   &  $f_0$         &  $b$    & $R_{opt} $ & $\beta$ \\
         &  (Kpc$^{-1}$)  &  (Kpc)  & (Kpc)      &         \\\hline
NGC1560  &  0.0726        &  2.119  & 4.6        & 0.344 \\
NGC2403  &  0.0171        &  5.399  & 6.7        & 0.546 \\
NGC3198  &  0.0038        &  12.88  & 11         & 0.547 \\
NGC6503  &  0.0290        &  3.035  & 3.8        & 0.702 \\
\end{tabular}   
\end{center}
\end{table}

\end{multicols}
\end{document}